# Dark Matter and Cosmic Acceleration from Wesson's IMT[*]

Mark Israelit[1]

In the present work a procedure is build up, that allows obtaining dark matter and cosmic acceleration in our 4D universe embedded in a 5D manifold. Both, dark matter and the factor causing cosmic acceleration, as well ordinary matter are induced in the 4D space-time by a warped, but empty from matter, 5D bulk. The procedure is carried out in the framework of the Weyl-Dirac version (Israelit, Found Phys 35:1725, 2005; Israelit, Found Phys 35:1769, 2005) of Paul Wesson's Induced Matter Theory (Wesson, Space-time matter, 1999) enriched by Rosen's approach (Found Phys 12:213, 1982). Considering chaotically oriented Weyl vector fields, which exist in microscopic cells, we obtain cold dark matter consisting of weylons, massive bosons having spin 1. Assuming homogeneity and isotropy at large scale we derive cosmological equations in which luminous matter, cold dark matter and dark energy may be considered separately. Making in the given procedure use of present observational data one can develop a model of the Universe with conventional matter, dark matter and cosmic acceleration, induced by the 5D bulk.

[1] Department of Physics and Mathematics, University of Haifa-Oranim, Tivon 36006 Israel.
  e-mail: <israelit@macam.ac.il>

# 1. INTRODUCTION

Two interesting phenomena are for the fast decades in the spotlight of cosmology: dark matter (DM) and the acceleration of the universe at present (cf. e.g. [1-3]) These remarkable cosmological events are verified observationally, but the origin of dark matter and of dark energy (DE) causing cosmic acceleration is till unknown.

At present it is believed that DM and DE make up ~ 96% of the mass/energy content of the Universe: DM contributes ~25%, DE - up to 71%. The latter affects the expansion rate of the Universe, which in turn affects the growth of structure and the distances to objects. It was found that the Universe decelerated until $z = 0.5$ and age 9Gyr, when it began accelerating and that the expansion has actually been speeding up for the past $5\text{Gyr} \approx 5 \times 10^9$ years [4-7].

There exist a number of theories trying to explain the origin of dark matter, but within the context of the general theory of relativity, cosmic acceleration cannot be explained by any form of conventional matter or energy. It can however be accommodated by a nearly smooth and very elastic ($P \approx -\rho$) form of DE that account for about 71% of the mass/energy content of the Universe.

Thus, what is the nature of DM, that holds the Universe together and of DE that is responsible for the expansion of the Universe to speed up? In the present paper an attempt is made to answer this question. Issuing from Wesson's Induced Matter Theory (MIT) [8] in its Weyl-Dirac (W-D) [9, 10] modification and applying Rosen's [11] approach, we have carried out a procedure, that allows getting luminous matter, DM and cosmic acceleration in our Universe, the latter regarded as a 4D brane in the 5D bulk.

Wesson's IMT is based on Einstein's general theory of relativity extended by the idea that our 4D space-time is a surface embedded in a 5-dimensional manifold, the bulk.

As long ago, in 1921, Kaluza proceeding from this idea proposed a unification of electromagnetism and gravitation [12] in the frame of a 4D hypersurface embedded in a 5-dimensional (5D) manifold. Suggesting that the fifth dimension has a circular topology Klein [13] imposed the cylindricity condition and completed the Kaluza theory. In the Kaluza-Klein theory, the fifth coordinate $x^4$ plays a purely formal role and the components of the 5D metric tensor do not depend on $x^4$. In order to obtain a better version of the unified theory of gravitation and electromagnetism A. Einstein and W. Mayer [14, 15] considered Kaluza's idea, from the standpoint that the space-time is a 4-dimensional one, however possessing vectors (and tensors) with a fifth component. In 1938 A. Einstein and P. G. Bergmann [16, 17] presented a generalization of the Kaluza-Klein theory. In this work the condition of cylindricity (that is equivalent to the existence of a 5D Killing vector) is replaced by the assumption that with regard to the fifth coordinate the space is periodically closed. In the Einstein - Bergmann version the fifth dimension has a physical meaning.

The Kaluza-Klein idea of extra dimensions, where matter is confined within a lower dimensional surface, has received an enormous amount of attention during the last decades. There must be noted the early works of Joseph [18], Akama [19], Rubakov and Shaposhnikov [20], Visser [21]. The basic works of Randall and Sundrum [22, 23], as well the works of Arkani-Hamed et al [24, 25], who suggested that ordinary matter would be confined to our 4D universe, while gravity would "live" in the extended 5-dimensional manifold, played a key role in the further development of Kaluza-Klein theories. .

On a revised Kaluza-Klein (K-K) approach is based Wesson's Induced Matter Theory (IMT) known also as the Space Time Matter Theory (STM). Among the originating works must be noted the papers of P. S. Wesson and J. Ponce de Leon [26-29]. Especially important is the work of Overduin and Wesson [30] as well the monograph by Wesson [8]. It was shown in these works that the cylindrical assumption may be eliminated and that the revised K-K theory is appropriate for dealing with cosmological problems and contains an induced electromagnetic field on the brane. Basic concepts and approaches of this remarkable theory as well applications to cosmology were developed during the last two decades in collaboration by Liko, Lim, Liu, Overduin, Ponce de Leon, Seahra and Wesson [8, 26-49]. Between important achievements of the Induced Matter Theory is the elegant proof of the geometric origin of matter. Regarding our 4D universe as a hypersurface (brane) embedded in a 5D space (bulk) Paul Wesson and collaborators have shown that the matter being in the 4D universe is induced by the *geometry* of the 5D bulk, the latter being a warped, but empty from matter one. Among works in the framework of the IMT are those dealing with test particles in higher-dimensional models, successful cosmological models, cosmological models with a variable cosmological constant, papers on dark matter, dark energy and induced unified theory of gravitation and electromagnetism.

Some years ago the present writer has shown that in Wesson's IMT, the 5D manifold (bulk) is rather a Weyl space, than a Riemannian one. The Weyl-Dirac [W-D] modification of Wesson's IMT was presented [9, 10]. Following the ideas of Weyl [50-52] and Dirac [53] in each point of the bulk, in addition to the metric tensor $g_{AB} = g_{BA}$, the Weyl connection vector $\tilde{w}_A$ and the Dirac gauge function $\Omega$ were introduced. Regarding the 5D Weyl field $\tilde{w}_A$ as the prototype of the 4D electromagnetic potential one obtains on the 4D brane both, gravitation and electromagnetism, as well gravitational matter and electric current. As $g_{AB}, \tilde{w}_A, \Omega$ are integral parts of the W-D geometry describing the bulk one concludes that the latter, being warped but empty, induces on the brane the unified theory. It must be pointed out that a unified theory of gravity and electromagnetism was already obtained in the original IMT [3, 30, 39] by a K-K procedure. Recently, in the framework of the W-D modification of Wesson's IMT, models of classical (non-quantum) neutral and charged fundamental particles induced by the bulk were presented [54, 55].

It turns out that the W-D bulk of Wesson's IMT can create DM and DE on the brane. As long ago, in 1982, Nathan Rosen [11] in discussing the (4-dimensional) Dirac modification of Weyl's theory pointed out that Dirac had chosen a particular value for a certain parameter appearing in his variational principle and that this value could be taken differently. Rosen showed that the parameter could be chosen so that, instead of an electromagnetic field, one gets a Proca [56] vector field which, from the standpoint of quantum mechanics, can be interpreted as an ensemble of particles having finite mass and spin 1. Later [57-60] it was suggested by Nathan Rosen and the present writer that these particles, named weylons, interacting gravitationally with ordinary matter, form the CDM in the universe. For the ensemble of weylons in the universe an acceptable cosmological behavior was obtained [58].

In order to get in the framework of Wesson's IMT a 4D brane with induced DM and cosmic acceleration, we will apply Rosen's approach [11] to the 5D W-D bulk.

**In the present work following conventions are valid.** Upper case Latin indices run from 0 to 4; lowercase Greek indices run from 0 to 3. Partial differentiation is denoted by a comma (,), Riemannian covariant 4D differentiation by a semicolon (;), and Riemannian covariant 5D differentiation by a colon (:). Further, the 5D metric tensor is denoted by $g_{AB}$, its 4D counterpart by $h_{\mu\nu}$; sometimes 5D quantities will be marked by a tilde, so $R_2^1$ is the component of the 4D Ricci tensor, whereas $\tilde{R}_2^1$ belongs to the 5D one, $R \equiv R_\sigma^\sigma$ is the 4D curvature scalar, $\tilde{R} \equiv \tilde{R}_S^S$ - the 5D one, $\tilde{w}_A$ stands for the 5D Weyl vector, while $w_\nu$ is its 4D counterpart.

2. THE 5D WEYL-DIRAC BULK AND THE 4D BRANE (Cf. [9, 46, 47])

In the original 5D IMT of Wesson, one regards the gravitational 5D bulk as pure geometry (gravitation) without any additional fields. The geometry is described by the metric tensor $g_{AB} = g_{BA}$. Thus, the principal phenomenon, which carries information, is a metric perturbation propagating in the form of a gravitational wave. In order to avoid misinterpretations one must assume that all gravitational waves have the same speed. Therefore, the isotropic 5-dimensional interval $dS^2 = 0$ has to be invariant, whereas an arbitrary line element $dS^2 = g_{AB}dx^A dx^B$ may vary. The situation resembles the 4D Weyl geometry [11, 50-52, 60] where the light cone is the principal phenomenon describing the space-time and hence the light-like interval $ds^2 = 0$ is invariant rather than an arbitrary line-element $ds^2 = h_{\alpha\beta}dy^\alpha dy^\beta$ between two space-time events. Following the ideas of Weyl [50-52] and Dirac [53] developed by Nathan Rosen [11] and the present writer [60] the Weyl-Dirac version of Wesson's IMT was proposed recently [9, 10]. In that version the 5D manifold {$M$} (the bulk) is mapped by coordinates $\{x^N\}$ and in every point exists the symmetric metric tensor $g_{AB}$, as well the Weylian connection vector $\tilde{w}_C$ and the Dirac gauge function $\Omega$.

The three fields $g_{AB}$, $\tilde{w}_C$ and $\Omega$ are integral parts of the geometric framework, and no additional fields or particles exist in the bulk {$M$}. In this 5D manifold, field equations, for $g_{AB}$, $\tilde{w}_C$, and $\Omega$ are derived from a geometrically based action.

Below, following the works of P. S. Wesson and S. S. Seahra [46, 47], a concise description of the general embedding formalism is given (Cf. [9, 55]).

One considers a 5-dimensional manifold { $M$ } (the "bulk") with a symmetric metric $g_{AB} = g_{BC}$, (A, B, =0, 1, 2, 3, 4) having the signature $\text{sig}(g_{AB}) = (+,-,-,-,\varepsilon)$ with $\varepsilon = \pm 1$. The manifold is mapped by coordinates { $x^A$ } and described by the line-element

$$dS^2 = g_{AB} dx^A dx^B \tag{1}$$

One can introduce a scalar function $l = l(x^A)$ that defines the foliation of {$M$} with 4-dimensional hyper-surfaces $\Sigma_l$ at a chosen $l$ = const, as well the vector $n^A$ normal to $\Sigma_l$. If there is only one time-like direction in {$M$}, it will be assumed that $n^A$ is space-like. If {$M$} possesses two time-like directions $(\varepsilon = +1)$, $n^A$ is a time-like vector. Thus, in any case $\Sigma_l$ (the "brane") contains three space-like directions and a time-like one. The brane, our 4-dimensional space-time, is mapped by coordinates { $y^\mu$ }, ($\mu = 0,1,2,3.$) and has the metric $h_{\mu\nu} = h_{\nu\mu}$ with $\text{sig}(h_{\mu\nu}) = (+,-,-,-)$. The line-element on the brane is

$$ds^2 = h_{\mu\nu} dy^\mu dy^\nu \tag{2}$$

It is supposed that the relations $y^\nu = y^\nu(x^A)$ and $l = l(x^A)$, as well as the reciprocal one $x^A = x^A(y^\nu, l)$ are mathematically well-behaved functions. Thus, the 5D bulk may be mapped either by $\{x^A\}$ or by $\{y^\nu, l\}$. One can consider a 5D normal vector to $\Sigma_l$

$$n_A = \varepsilon \Phi \partial_A l; \tag{3}$$

with $\Phi$ being the lapse function.

A 5D quantity (vector, tensor) in the bulk has 4D counterparts located on the brane. These counterparts may be formed by means of the following system of basis vectors, which are orthogonal to $n_A$

$$e_v^A = \frac{\partial x^A}{\partial y^v} \quad \text{with} \quad n_A e_v^A = 0 \tag{4}$$

The brane $\Sigma_l$ is stretched on four $(v = 0,1,2,3)$ five-dimensional basis vectors $e_v^A$. Together with the main basis $\{e_v^A; n_A\}$ one can consider its associated one $\{e_A^v; n^A\}$, which also satisfies the orthogonality condition $e_A^v n^A = 0$. The main basis and its associated are connected by the following relations:

$$e_v^A e_A^\mu = \delta_v^\mu \; ; \quad e_\sigma^A e_B^\sigma = \delta_B^A - \varepsilon n^A n_B \; ; \quad n^A n_A = \varepsilon \tag{5}$$

Let us consider a 5D vector $V_A; V^A$ in the bulk $\{M\}$. Its 4D counterpart on the brane $\Sigma_l$ is given by

$$V_\mu = e_\mu^A V_A \; ; \quad V^v = e_B^v V^B . \tag{6}$$

On the other hand the 5D vector may be written as

$$V_A = e_A^\mu V_\mu + \varepsilon (V_S n^S) n_A ; \quad V^A = e_\mu^A V^\mu + \varepsilon (V^S n_S) n^A \tag{7}$$

Actually, (7) is a decomposition of $V_A$ into a 4-vector $V_\mu$ and a part normal to $\Sigma_l$. Further, the 5D metric tensor, $g_{AB}; g^{AB}$ and the 4D one $h_{\mu\nu}; h^{\mu\nu}$ are related by

$$h_{\mu\nu} = e_\mu^A e_\nu^B g_{AB} \; ; \quad h^{\mu\nu} = e_A^\mu e_B^\nu g^{AB} \; ; \quad \text{with} \quad h_{\mu\nu} h^{\lambda\nu} = \delta_\mu^\lambda \tag{8}$$

and

$$g_{AB} = e_A^\mu e_B^\nu h_{\mu\nu} + \varepsilon n_A n_B \; ; \quad g^{AB} = e_\mu^A e_\nu^B h^{\mu\nu} + \varepsilon n^A n^B \; ; \quad \text{with} \quad g_{AB} g^{CB} = \delta_A^C \tag{9}$$

## 3. THE ACTION INTEGRAL AND THE FIELD EQUATIONS IN THE BULK

Being guided by Dirac's approach [53] we derive the field equations in the 5D bulk from a variational principle $\delta \int L_{geom} \sqrt{-g} \, d^5 x = 0$, with $L_{geom} \sqrt{-g}$ being an in-invariant, i.e. invariant under both, coordinate transformations and Weyl gauge transformations. The Lagrangian $L_{geom}$ is formed from a number of suitable, geometrically based terms. Making use of $g_{AB}$, $\tilde{w}_C$, $\Omega$ and their derivatives we build the following action integral (cf. [9])

$$I = \int \left[ \Omega \tilde{W}^{AB} \tilde{W}_{AB} - \Omega^3 \tilde{R} + \sigma \Omega^2 \tilde{w}^L (\Omega \tilde{w}_L + 2\Omega_L) + (\sigma + 12)\Omega \Omega^L \Omega_L + \Omega^5 \Lambda \right] \sqrt{-g} \, d^5 x \quad (10)$$

In the action (10) appear the following quantities: the Weyl length curvature tensor $\tilde{W}_{AB} \equiv \tilde{w}_{A,B} - \tilde{w}_{B,A}$; the 5D Riemannian curvature scalar $\tilde{R} \equiv \tilde{R}_S^S$, formed from the 5D Christoffel symbols $\tilde{\Gamma}_{AB}^C$; Dirac's gauge function $\Omega$ and its partial derivatives $\Omega_A \equiv \Omega_{,A} = \frac{\partial \Omega}{\partial x^A}$, $\Omega^A = g^{AB} \Omega_B$. The symbol $\Lambda$ stands for the cosmological constant. Finally, $\sigma$ is an arbitrary constant. The dynamical variables in (10) are $g_{AB}$, $\tilde{w}_D$ and $\Omega$.

In the present work we want to get on the 4D brane dark energy causing cosmic acceleration as well dark matter, both induced by the bulk. Therefore we adopt Rosen's [11] idea and assume that $\sigma \neq 0$; and that $\sigma$ can obtain arbitrary values[2]. In this case the Weyl field ($\tilde{w}_A$; $\tilde{W}_{AB}$) of the bulk is no longer the 5D creator of the 4D Maxwell field.

Considering the variation of (10) with respect to the dynamical variables $g_{AB}$, $w_D$ and $\Omega$, one obtains the field equations in the bulk.

Varying $g^{AB}$ one obtains the "gravitational" equation

$$\tilde{G}_{AB} = -\frac{8\pi}{\Omega^2} \tilde{M}_{AB} + \sigma \left( \tilde{w}_A \tilde{w}_B - \frac{1}{2} g_{AB} \tilde{w}^L \tilde{w}_L \right) + \frac{\sigma}{\Omega} \left( 2\Omega_A \tilde{w}_B - g_{AB} \Omega^L \tilde{w}_L \right) +$$
$$+ (\sigma + 6) \frac{\Omega_A \Omega_B}{\Omega \Omega} - \frac{1}{2} \sigma g_{AB} \frac{\Omega^L \Omega_L}{\Omega \Omega} + \frac{3}{\Omega} \left( g_{AB} \Omega^L_{:L} - \Omega_{A:B} \right) - \frac{1}{2} g_{AB} \Omega^2 \Lambda \quad (11)$$

Here $\tilde{G}_{AB} \equiv \tilde{R}_{AB} - \frac{1}{2} g_{AB} \tilde{R}$ is the 5D Einstein tensor, and

$$4\pi \tilde{M}_{AB} \equiv \left(\frac{1}{4}g_{AB}\tilde{W}^{LN}\tilde{W}_{LN} - \tilde{W}_{AL}\tilde{W}_B{}^L\right) \quad (12)$$

---

[2] Setting $\sigma = 0$ one would obtain from (10) the action integral of the W-D modification considered in [1], so that on the brane one would have a geometrically based unified theory of gravitation and electromagnetism with mass and current induced by the bulk.

$\tilde{M}_{AB}$ can be regarded as the energy-momentum density tensor of the 5D Weyl field.

Varying $\tilde{w}_A$ one finds

$$\left(\Omega\tilde{W}^{AB}\right)_{:B} = \frac{1}{2}\sigma\Omega^2 g^{LA}\left(\Omega\tilde{w}_L + \Omega_L\right) \quad (13)$$

Let us consider the $\Omega$-equation. On one hand varying $\Omega$ in (10) we obtain

$$3\Omega^2 \tilde{R} = \tilde{W}^{AB}\tilde{W}_{AB} + 5\Omega^4\Lambda + \sigma\Omega^2\left(3\tilde{w}^L\tilde{w}_L - 2\tilde{w}^L_{:L}\right) - (\sigma+12)\left(\Omega^L\Omega_L + 2\Omega\Omega^L_{:L}\right) \quad (14)$$

On the other hand, contracting (11) we have

$$3\tilde{R} = \frac{1}{\Omega^2}\tilde{W}^{AB}\tilde{W}_{AB} + 3\sigma\,\tilde{w}^A\tilde{w}_A + 6\sigma\frac{\Omega_L}{\Omega}\tilde{w}^L - (12 - 3\sigma)\frac{\Omega^L}{\Omega}\frac{\Omega_L}{\Omega} - 24\frac{\Omega^L_{:L}}{\Omega} + 5\Omega^2\Lambda \quad (15).$$

Comparing (14) and (15) we obtain

$$\sigma\left(\tilde{w}^L_{:L} + 3\frac{\Omega_L}{\Omega}\tilde{w}^L + \frac{\Omega^L_{:L}}{\Omega} + 2\frac{\Omega^L}{\Omega}\frac{\Omega_L}{\Omega}\right) = 0 \quad (16)$$

It seems that EQ. (16) is a restriction on $\Omega$. However, turning to equation (13), one obtains $\left[\sigma\Omega^2 g^{LA}\left(\Omega\tilde{w}_L + \Omega_L\right)\right]_{:A} = 0$, and this satisfies (16). Thus, the $\Omega$-equation (14) is a corollary of (11) and (13), and for $\sigma \neq 0$ (as it was for $\sigma = 0$) the Dirac gauge function may be chosen ***arbitrarily.*** (cf. [9]; [60])

Considering cosmological models we believe that at large scale the universe is isotropic and homogeneous and no cosmic vector functions are present. However, we can adopt the existence of microscopic areas with chaotically oriented vector fields. This standpoint will be applied to the 5D Weyl-Dirac bulk. Thus, on the one hand the bulk has a microstructure caused by chaotically oriented local Weyl vector fields $\tilde{w}_A{}_{loc}$, on the other hand at large cosmic scales (globally) the bulk is described by scalar fields.

Let us first consider the situation at large scales. From EQ. (13) we see that setting $\tilde{w}_A = 0$ would imply either $\sigma = 0$ or $\Omega = \text{const}$. We can, however, assume that $\tilde{w}_A{}_{global}$ is a 5D gradient vector

$$\tilde{w}_A{}_{global} = -\frac{\Omega_A}{\Omega} \quad (17)$$

By (17), EQ. (13) is satisfied identically and in EQ. (11) are no vector fields.

One can describe both, local and global effects if one writes

$$\tilde{w}_A = \underset{loc}{\tilde{w}_A} + \underset{global}{\tilde{w}_A} = \underset{loc}{\tilde{w}_A} - \frac{\Omega_A}{\Omega} \tag{18}$$

With the separation (18) between the global and local fields, one rewrites EQ. (11) as

$$\tilde{G}_{AB} = -\frac{8\pi}{\Omega^2} \underset{loc}{\tilde{M}_{AB}} + \sigma\left(\underset{loc}{\tilde{w}_A} \underset{loc}{\tilde{w}_B} - \frac{1}{2} g_{AB} \underset{loc}{\tilde{w}^L} \underset{loc}{\tilde{w}_L}\right) + 6\frac{\Omega_A}{\Omega}\frac{\Omega_B}{\Omega} + \frac{3}{\Omega}\left(g_{AB}\Omega^L_{:L} - \Omega_{A:B}\right) - \frac{1}{2} g_{AB}\Omega^2 \Lambda \tag{19}$$

and EQ. (13) as

$$\left(\Omega \underset{loc}{\tilde{W}}^{AB}\right)_{:B} = \frac{1}{2}\sigma \Omega^3 \underset{loc}{\tilde{w}^A} \tag{20}$$

At large scales, where $\underset{loc}{\tilde{w}_A}$ is irrelevant, we obtain from EQ. (19)

$$\tilde{G}_{AB} = -\frac{8\pi}{\Omega^2} \underset{global}{\tilde{T}_{AB}} + 6\frac{\Omega_A}{\Omega}\frac{\Omega_B}{\Omega} + \frac{3}{\Omega}\left(g_{AB}\Omega^L_{:L} - \Omega_{A:B}\right) - \frac{1}{2} g_{AB}\Omega^2 \Lambda \tag{19a}$$

In (19a) $\underset{global}{\tilde{T}_{AB}}$ represent the global effect caused by the local cells of the microstructure.

Turning to the local effect we point out that in a 5D very small micro-region, the Dirac gauge function $\Omega$ remains nearly unchanged, so that inside the micro-region one can take the Einstein gauge $\Omega = 1$. Then EQ. (19) takes the form

$$\tilde{G}_{AB} = -8\pi \underset{loc}{\tilde{M}_{AB}} + \sigma\left(\underset{loc}{\tilde{w}_A} \underset{loc}{\tilde{w}_B} - \frac{1}{2} g_{AB} \underset{loc}{\tilde{w}^L} \underset{loc}{\tilde{w}_L}\right) - \frac{1}{2} g_{AB}\Lambda \tag{19b}$$

whereas EQ. (20) may be written as

$$\left(\underset{loc}{\tilde{W}}^{AB}\right)_{:B} = \frac{1}{2}\sigma \underset{loc}{\tilde{w}^A} \tag{20a}$$

The local field on the brane will be considered in detail in section 5.

## 4. EQUATIONS ON THE 4D BRANE

There are two sets of equations in the brane; the first, induced by EQ. (11), gives the equations of gravitation, the second set, being derived from EQ. (13), describes the Weylian field.

The equations of gravitation may be obtained by inserting EQ. (11) into the Gauss equations (cf. [9, 46, 47])

$$R_{\alpha\beta} = e^A_\alpha e^B_\beta \tilde{R}_{AB} + \varepsilon\left[E_{\alpha\beta} - 2h^{\lambda\sigma} C_{\lambda[\sigma} C_{\beta]\alpha}\right] \qquad R = \tilde{R} + 2\varepsilon\left[E - h^{\lambda\sigma} h^{\mu\nu} C_{\mu[\nu} C_{\lambda]\sigma}\right] \qquad (21)$$

In (21) appear the basis vectors $e^A_\alpha$ (cf. (4), (5)), as well the extrinsic curvature of the brane embedded in the bulk

$$C_{\mu\nu} = e^A_\mu e^B_\nu \, n_{B:A} \equiv e^A_\mu e^B_\nu \left(\frac{\partial n_B}{\partial x^A} - n_S \tilde{\Gamma}^S_{AB}\right) \qquad (21a)$$

There is also the quantity

$$E_{\alpha\beta} \equiv \tilde{R}_{MANB}\, n^M n^N e^A_\alpha e^B_\beta \;, \text{ and its contraction } E \equiv h^{\alpha\beta} E_{\alpha\varepsilon\beta} = -\tilde{R}_{MN}\, n^M n^N \qquad (21b)$$

The expressions (21b) present induced by the warped bulk matter. Calculating $\tilde{R}_{AB}$ and $\tilde{R}$ from EQ. (11) and substituting into (21) one obtains the 4D gravitational equation

$$\begin{aligned}
G_{\alpha\beta} &= -\frac{8\pi}{\Omega^2} M_{\alpha\beta} - \frac{2\varepsilon}{\Omega^2}\left(\frac{1}{2} h_{\alpha\beta} B - B_{\alpha\beta}\right) + \frac{3}{\Omega^2}\left(2\Omega_\alpha \Omega_\beta - \Omega\Omega_{\alpha;\beta} + h_{\alpha\beta}\Omega\Omega^\sigma_{;\sigma}\right) + \\
&+ \frac{3\varepsilon}{\Omega}\left(\Omega^S n_S\right)\left(h_{\alpha\beta} C - C_{\alpha\beta}\right) + \varepsilon\left[E_{\alpha\beta} - h_{\alpha\beta} E + h^{\mu\nu} C_{\mu[\nu} C_{\lambda]\sigma}\left(h_{\alpha\beta} h^{\lambda\sigma} - 2\delta^\sigma_\alpha \delta^\lambda_\beta\right)\right] + \\
&+ \sigma\left[\left(w_\alpha w_\beta - \frac{h_{\alpha\beta}}{2} w_\sigma w^\sigma\right) + \frac{2}{\Omega}\left(w_\alpha \Omega_\beta - \frac{h_{\alpha\beta}}{2} w_\sigma \Omega^\sigma\right) + \frac{1}{\Omega^2}\left(\Omega_\alpha \Omega_\beta - \frac{h_{\alpha\beta}}{2}\Omega_\sigma \Omega^\sigma\right)\right] - \\
&- \sigma\varepsilon \frac{h_{\alpha\beta}}{2}\left[\left(n_S w^S\right)^2 + \frac{1}{\Omega^2}\left(n_S \Omega^S\right)^2 + \frac{2}{\Omega}\left(n_S w^S\right)\left(n^R \Omega_R\right)\right] - \frac{1}{2} h_{\alpha\beta}\Omega^2 \Lambda
\end{aligned} \qquad (22)$$

In EQ. (22) appear the quantities introduced in (21a, b)), as well $C \equiv h^{\lambda\sigma} C_{\lambda\sigma}$. There are also the quantities

$$B_{\alpha\beta} \equiv \tilde{W}_{AS} \tilde{W}_{BL} e^A_\alpha e^B_\beta n^S n^L \;; \qquad \text{and} \qquad B = h^{\lambda\sigma} B_{\lambda\sigma} \equiv \tilde{W}_{AS} \tilde{W}_{BL} g^{AB} n^S n^L \;, \qquad (22a),$$

which are induced by energy-like expressions of the 5D Weyl field. Finally, in (22) appears the expression $M_{\alpha\beta} \equiv \frac{1}{4\pi}\left(\frac{1}{4}h_{\alpha\beta}W^{\lambda\sigma}W_{\lambda\sigma} - W_{\alpha\lambda}W_\beta{}^\lambda\right)$, which is the energy-momentum density tensor of the 4D Weyl field.

From (13) one obtains the equation for the Weylian field on the brane

$$W^{\alpha\beta}_{;\beta} = -\frac{\Omega_\beta}{\Omega}W^{\alpha\beta} - \varepsilon\, n_S\left[e^\alpha_A\, n^C\left(\tilde{W}^{AS}_{:C} + \frac{\Omega_C}{\Omega}\tilde{W}^{AS}\right) + \tilde{W}^{AS}\left(e^\alpha_A h^{\beta\lambda} - e^\beta_A h^{\alpha\lambda}\right)C_{\beta\lambda}\right] + \tag{23}$$
$$+ \frac{1}{2}\sigma\Omega^2\left(w^\alpha + \frac{\Omega^\alpha}{\Omega}\right)$$

It is worth noting that $\sigma \neq 0$, so that in the present framework $\tilde{W}_{AB}$ is **not** the prototype of a 4D Maxwell field, as it was in previous papers [9, 54, 55].

As pointed out above, we believe that at large scale the universe is isotropic and homogeneous and no cosmic vector functions are present. However, we allow the existence of small areas with chaotically oriented vector fields. This standpoint was applied to the 5D bulk, and by (6) it is valid in our universe, the 4D brane. According to (6) and (18) the Weyl field on the brane, which is induced by its 5D counterpart, is divided into a cosmic gradient vector field and a chaotically oriented local vector field

$$\underset{\text{total}}{w_\nu} = e^A_\nu \tilde{w}_A = e^A_\nu\left(\underset{\text{loc}}{\tilde{w}_A} - \frac{\Omega_A}{\Omega}\right) = \left(\underset{\text{loc}}{w_\nu} + \underset{\text{glob}}{w_\nu}\right) = w_\nu - \frac{\Omega_\nu}{\Omega} \tag{24}$$

Omitting the subscribe "$_{loc}$" we denote from now on by $w_\nu$ the local chaotically oriented Weyl vector field in microscopic areas.

## 5. LOCAL WEYL FIELDS ON THE BRANE

One turns to a very small 4D micro-region, a cell, in which one can take the Einstein gauge $\Omega = 1$, (cf (19b), (20a)), so that from (22) follows

$$G_{\alpha\beta} = -8\pi M_{\alpha\beta} + \sigma\left(w_\alpha w_\beta - \frac{h_{\alpha\beta}}{2}w_\sigma w^\sigma\right) - -\frac{1}{2}h_{\alpha\beta}\Lambda - \sigma\,\varepsilon\frac{h_{\alpha\beta}}{2}(n_S w^S)^2$$
$$+ \varepsilon\left[E_{\alpha\beta} - h_{\alpha\beta}E + h^{\mu\nu}C_{\mu[\nu}C_{\lambda]\sigma}\left(h_{\alpha\beta}h^{\lambda\sigma} - 2\delta^\sigma_\alpha\delta^\lambda_\beta\right)\right] - 2\varepsilon\left(\frac{1}{2}h_{\alpha\beta}B - B_{\alpha\beta}\right) \tag{25}$$

and from (23) one obtains

$$W^{\alpha\beta}_{;\beta} = \frac{1}{2}\sigma w^\alpha + \varepsilon n_S \left[ \tilde{W}^{LS} \left( e^\gamma_L h^{\alpha\lambda} - e^\alpha_L h^{\gamma\lambda} \right) C_{\lambda\gamma} - n^C e^\alpha_A \tilde{W}^{AS}_{:C} \right] \tag{26}$$

Considering an empty micro-cell, with no induced matter and current (the $\varepsilon-$ terms in equations (25, 26)), and discarding also the cosmological term in (25) we obtain

$$G_{\alpha\beta} = -8\pi M_{\alpha\beta} + \sigma\left( w_\alpha w_\beta - \frac{h_{\alpha\beta}}{2} w_\sigma w^\sigma \right) \tag{27}$$

and the Proca EQ. [56]

$$W^{\alpha\beta}_{;\beta} = \frac{1}{2}\sigma w^\alpha \tag{28}$$

EQ. (27), with the RHS being the energy-momentum density of the Weylian field inside the cell, may be rewritten explicitly as

$$G^\beta_\alpha = -8\pi \, T_{\text{cell}} = 2W_{\alpha\sigma}W^{\beta\sigma} - \frac{1}{2}\delta^\beta_\alpha W_{\lambda\sigma}W^{\lambda\sigma} + \frac{1}{2}\sigma\left(2w_\alpha w^\beta - \delta^\beta_\alpha w_\sigma w^\sigma\right) \tag{29}$$

Turning to (28) one obtains a Lorentz-like condition

$$w^\mu_{;\mu} = 0 \tag{30}$$

Taking into account the relation $W_{\mu\nu} = w_{\mu,\nu} - w_{\nu,\mu}$ as well (30), one can rewrite (28) as

$$h^{\alpha\beta} w^\mu_{;\alpha;\beta} + w^\alpha R^\mu_\alpha - \frac{1}{2}\sigma w^\mu = 0 \tag{31}$$

Finally, neglecting inside the micro-cell the curvature, choosing negative values of $\sigma$ and writing $\sigma = -2\kappa^2$, we obtain

$$h^{\alpha\beta} w^\mu_{;\alpha;\beta} + \kappa^2 w^\mu = 0 \tag{32}$$

This is the equation for a vector boson field of particles having spin 1 and mass $m$, the latter according to quantum mechanics being linked to $\kappa$ (in conventional units) by

$$\kappa = \frac{mc}{\hbar} = \frac{2\pi}{\lambda_C}; \text{ so that } m = \frac{\hbar\kappa}{c} \tag{33}$$

In (33) $\lambda_C$ is the Compton wavelength. As ordinary matter has no Weyl charge [cf. 11, 60] the boson, named weylon, interacts with ordinary luminous matter only trough gravitation and not directly.

As the vector fields inside the cells are chaotically oriented, the summary angular momentum in a macroscopic region vanishes, and we have only a mass effect caused by the particles. Consequently, we can consider the ensemble of these bosons as dark matter and develop an appropriate cosmological framework.

It must, however, be noted that 16 years ago in a paper [57] by Nathan Rosen and the present writer, a 4-dimensional Weyl-Dirac scenario was developed in which bosons, named weylons, constitute dark matter in the universe. It was proposed that our universe, being filled with luminous matter, has a chaotic Weylian microstructure, but is described on large scale by the F-R-W metric. The Weylian microstructure creates particles-weylons, that interact with ordinary matter only trough gravitation and not directly, so that weylons are analogous to photons and gravitons but differ in that they are massive. Two years later the same authors [58] have investigated the behavior of Weylian dark matter in the universe, described either by the non-singular model [61] or by the standard one [62]. It was shown that the ensemble of weylons, being described by the Bose-Einstein statistics, may be considered in the universe as a gas, the thermal behavior of which is given by means of the Boltzmann statistics, provided the weylon mass, $m_W \gg 4.1 \text{eV}$. The behavior of this DM was considered in [58] from the time when weylons were created and until the present. For weylons having mass $10 \text{MeV} \leq m_W \leq 10^5 \text{GeV}$ an acceptable cosmological behavior was obtained. It was also shown in [58] that this weylon DM was unimportant in the early stages of the universe, but became important at the time of galaxy formation and may have played a role in this process.

One should distinguish between the DM in and around cosmic inhomogeneities (like galaxies, clusters, etc.) and the DM pervading all of the cosmic space [59]. The first one, known as Cold Dark Matter (CDM), may be presented by the weylon gas considered above.

***As the assumptions concerning weylons as well equations* (cf. (29), (32))** ***of the present work coincide with assumptions and equations that were presented in*** [57]***, we will adopt the main results and conclusions of*** [57, 58] ***without repeatedly developing the formalism and proving the formulae.***

## 6. THE GLOBAL FIELD ON THE BRANE

In the previous section the local Weyl field as creator of cold dark matter (CDM) was considered. Now, let us turn to the gravitational equations of the global field in a FRW universe. Making use of the separation (24) we find that (23) is satisfied identically by $w_{total} = w_{\nu\, glob} = -\dfrac{\Omega_\nu}{\Omega}$ whereas from (22) we obtain the gravitational equation for the global field

$$G_\alpha^\beta = -\frac{8\pi}{\Omega^2} T_{\alpha\,CDM}^\beta + \frac{3}{\Omega^2}\left(2\Omega_\alpha \Omega^\beta - h^{\beta\lambda}\Omega\Omega_{\alpha;\lambda} + \delta_\alpha^\beta \Omega\Omega^\sigma_{;\sigma}\right) + \frac{3\varepsilon}{\Omega}\left(\Omega^S n_S\right)\left(\delta_\alpha^\beta C - C_\alpha^\beta\right)$$
$$+ \varepsilon\left[E_\alpha^\beta - \delta_\alpha^\beta E + h^{\mu\nu}C_{\mu[\nu}C_{\lambda]\sigma}\left(\delta_\alpha^\beta h^{\lambda\sigma} - 2\delta_\alpha^\sigma h^{\lambda\beta}\right)\right] - \frac{1}{2}\delta_\alpha^\beta \Omega^2 \Lambda \qquad (34)$$

In EQ. (34) the term $T_{\alpha\,CDM}^\beta$ represents the contribution of the CDM weylon gas, further $\Omega_\alpha \equiv \dfrac{\partial \Omega}{\partial y^\alpha}$, $\Omega^\alpha \equiv h^{\alpha\lambda}\Omega_\lambda$, and finally, the quantities $C_{\mu\nu}; C; E_\alpha^\beta \equiv h^{\beta\lambda}E_{\alpha\lambda}; E$ are defined above in (21a,b).

The bulk is mapped by coordinates $x^{0,1,2,3,4} \equiv t, r, \vartheta, \varphi, l$; and described by the line–element

$$dS^2 = dt^2 - R^2(t, l)\left[\frac{1}{1-kr^2}dr^2 + r^2\left(d\vartheta^2 + \sin^2\vartheta\, d\varphi^2\right)\right] + \varepsilon\,\Phi^2(t,l)dl^2 \qquad (35)$$

Our 4D universe, located on the brane $l = l_0$, is mapped by $y^{0,1,2,3} \equiv t, r, \vartheta, \varphi$ and described by the usual FRW line-element

$$ds^2 = dt^2 - R^2(t, l_0)\left[\frac{1}{1-kr^2}dr^2 + r^2\left(d\vartheta^2 + \sin^2\vartheta\, d\varphi^2\right)\right] \qquad (36)$$

with $k = 0, \pm 1$. According to (35), (36), the metric coefficients are[3]

$$g_{00} = 1;\ g_{11} = -\frac{R^2(t,l)}{1-kr^2};\ g_{22} = -R^2 r^2;\ g_{33} = g_{22}\sin^2\vartheta;\ g_{44} = \varepsilon\,\Phi^2(t,l);$$
$$h_{00} = 1;\ h_{11} = g_{11}(t,l_0);\ h_{22} = g_{22}(t,l_0);\ h_{33} = h_{22}\sin^2\vartheta; \qquad (37)$$

---

[3] A similar Ansatz was recently used by B. Mashhoon and P. Wesson [49].

In order to calculate $C_{\mu\nu}$; $C$; $E_\alpha^\beta \equiv h^{\beta\lambda} E_{\alpha\lambda}$; $E$ we need the basic vectors (cf. (4)

$$e_\alpha^A = \begin{matrix} 1 & 0 & 0 & 0 & 0 \\ 0 & 1 & 0 & 0 & 0 \\ 0 & 0 & 1 & 0 & 0 \\ 0 & 0 & 0 & 1 & 0 \end{matrix} \qquad e_A^\alpha = \begin{matrix} 1 & 0 & 0 & 0 & 0 \\ 0 & 1 & 0 & 0 & 0 \\ 0 & 0 & 1 & 0 & 0 \\ 0 & 0 & 0 & 1 & 0 \end{matrix}$$

(37a)

$$n_A = 0,0,0,0, -\varepsilon\, \Phi(t,l) \qquad n^A = 0,0,0,0, -\frac{1}{\Phi(t,l)}$$

Taking the homogeneous and isotropic universe to be filled with matter characterized by density $\rho(t) = T_0^0$ and pressure $P(t) = -T_1^1 = -T_2^2 = -T_3^3$, and making use of (21a, b) as well of (37), (37a), and denoting, $\partial_t f \equiv \dot{f}$; $\partial_l f \equiv f_{,4} \equiv f'$ we can write down the gravitational equations (34) explicitly

$$G_0^0 \equiv -3\left(\frac{(\dot{R})^2}{R^2} + \frac{k}{R^2}\right) = -8\pi \underset{\text{Total}}{\rho} = -\frac{8\pi}{\Omega^2} \underset{CDM}{T_0^0} + 3\left[2\frac{(\dot{\Omega})^2}{\Omega^2} + 3\frac{\dot{R}}{R}\frac{\dot{\Omega}}{\Omega} + \frac{\dot{R}\dot{\Phi}}{R\Phi}\right] +$$
$$+ 3\varepsilon\frac{1}{\Phi^2}\left[\frac{R''}{R} - \frac{R'}{R}\frac{\Phi'}{\Phi} + 3\frac{\Omega'}{\Omega}\frac{R'}{R}\right] - \frac{1}{2}\Omega^2 \Lambda$$

(38)

$$G_i^i \equiv -\left(2\frac{\ddot{R}}{R} + \frac{(\dot{R})^2}{R^2} + \frac{k}{R^2}\right) = 8\pi \underset{\text{Total}}{P} = -\frac{8\pi}{\Omega^2} \underset{CDM}{T_i^i} + \left[6\frac{\dot{R}\dot{\Omega}}{R\Omega} + 3\frac{\ddot{\Omega}}{\Omega} + 2\frac{\dot{R}\dot{\Phi}}{R\Phi} + \frac{\ddot{\Phi}}{\Phi}\right] +$$
$$+ 3\varepsilon\frac{1}{\Phi^2}\left[\frac{R''}{R} - \frac{R'\Phi'}{R\Phi} + 3\frac{R'\Omega'}{R\Omega}\right] - \frac{1}{2}\Omega^2 \Lambda; \qquad (i=1,2,3; \text{do not sum over } i)$$

(39)

It is worth noting that the terms in the second line of (38) and (39) describe a substance with the equation of state, $\rho + P = 0$ i.e. prematter (cf. [61]). From these terms it is possible to obtain a repulsing action that may lead to cosmic acceleration, alternatively, setting $\Lambda = 0$, one gets a cosmological constant induced by the bulk (Cf. the paper by Liu and Wesson, [40]).

In EQ.-s (38, 39), the gauge function, $\Omega(t, l)$ is an arbitrary but positive function, whereas $R(t, l)$ and $\Phi(t, l)$ are solutions of the gravitational equations.

To simplify the procedure one can turn to ordinary differential EQ.-s, assuming that the dependence of the function, $R, \Phi,$ and $\Omega$ on $t$ and $l$ may be separated as follows

$$R(t, l) = a(t)L(l); \quad \Phi(t, l) = \varphi(t)F(l) \quad \Omega(t, l) = \beta(t)O(l); . \qquad (40)$$

In addition, on the brane $l_0$ one can take $L(l_0) = F(l_0) == O(l_0) = 1$. (40a)

Inserting the separation condition (40) into (38, 39) we obtain

$$\left(\frac{(\dot{a})^2}{a^2} + \frac{k}{a^2}\right) = \frac{8\pi}{3}\rho_{Total} = \frac{8\pi}{3}\frac{1}{\beta^2}T^0_{0\,CDM} - \frac{\dot{a}\dot{\varphi}}{a\varphi} - \frac{1}{\varphi^2}B - \left[2\frac{(\dot{\beta})^2}{\beta^2} + 3\frac{\dot{a}}{a}\frac{\dot{\beta}}{\beta}\right] + \frac{1}{6}\beta^2\Lambda \qquad (41)$$

$$\frac{\ddot{a}}{a} = -\frac{4\pi}{3}\left(3P_{Total} + \rho_{Total}\right) = \frac{4\pi}{\beta^2}\left(T^i_{i\,CDM} - \frac{1}{3}T^0_{0\,CDM}\right) - \frac{1}{2}\left(\frac{\ddot{\varphi}}{\varphi} + \frac{\dot{a}\dot{\varphi}}{a\varphi}\right) - \frac{1}{\varphi^2}B +$$
$$+ \left[\frac{(\dot{\beta})^2}{\beta^2} - \frac{3}{2}\frac{\dot{a}\dot{\beta}}{a\beta} - \frac{3}{2}\frac{\ddot{\beta}}{\beta}\right] + \frac{1}{6}\beta^2\Lambda \qquad (42)$$

Here $a(t)$ is the 4D cosmic scale parameter, $\beta$ represents the 4D Dirac gauge function, which may be chosen arbitrarily, while the $\varphi$-terms describe the matter induced by the fifth coordinate of the bulk. There is also the induced by the bulk expression, $B \equiv \varepsilon\left[\dfrac{L''}{L} - \dfrac{L'}{L}\dfrac{F'}{F} + 3\dfrac{O'}{O}\dfrac{L'}{L}\right]$ which on the brane $l = l_0$ is obviously constant.

In EQ.-s (41, 42) we recognize the following components:

a) The density and pressure of induced by the bulk "ordinary", luminous matter

$$8\pi\rho = -3\frac{\dot{a}\dot{\varphi}}{a\varphi} - \frac{3}{\varphi^2}B \; ; \quad 8\pi P = 2\frac{\dot{a}\dot{\varphi}}{a\varphi} + \frac{\ddot{\varphi}}{\varphi} + 3\frac{1}{\varphi^2}B \qquad (43)$$

b) The effective density and pressure of the CDM weylon gas

$$8\pi\,\tilde{\rho}_{CDM} = 8\pi\frac{1}{\beta^2}T^0_{0\,CDM} \; ; \quad 8\pi\,\tilde{P}_{CDM} = -8\pi\frac{1}{\beta^2}T^1_{1\,CDM} \qquad (44)$$

c) The density and pressure induced by the Dirac gauge function $\beta$

$$8\pi \rho_{Beta} = -3\left[2\frac{(\dot{\beta})^2}{\beta^2} + 3\frac{\dot{a}}{a}\frac{\dot{\beta}}{\beta}\right] + \frac{1}{2}\beta^2\Lambda \ ; \quad 8\pi P_{Beta} = 3\left[2\frac{\dot{a}\dot{\beta}}{a\beta} + \frac{\ddot{\beta}}{\beta}\right] - \frac{1}{2}\beta^2\Lambda \qquad (45)$$

Thus, the total matter density and pressure my be written as

$$\rho_{Total} = \rho + \tilde{\rho}_{CDM} + \rho_{Beta} \ ; \qquad P_{Total} = P + \tilde{P}_{CDM} + P_{Beta} \ . \qquad (46)$$

It is believed that the $\beta-$matter (45) will form DE being responsible for the present cosmic acceleration.

The above considered procedure is supplying a plausible framework for constructing cosmological models, which contain induced by the 5D bulk luminous matter, cold dark matter and dark energy, the latter causing the present cosmic acceleration. It is worth emphasizing that all kinds of matter, the luminous, CDM and DE are induced by the 5D bulk in the framework of the W-D version [1, 2] of Wesson's geometrically based IMT [8, 26-49].

## 7. CONCLUSION and DISCUSSION

In this work we have made an attempt to explain the origin of DM and DE causing cosmic acceleration. In our scenario, all forms of matter, luminous, DM and DE are induced by a 5D bulk in the 4D brane, our Universe. The procedure is built up in the framework of the Wesson's IMT [8, 26-49], in its W-D version [9, 10], where the bulk is characterized rather by the Weyl-Dirac geometry [50-53], than by Riemann's. In addition an approach proposed by Nathan Rosen [11] is applied in order to get DM and DE..

Up to the observation of DM effects (the flat rotation curves of spiral galaxies and other evidences of "missing mass" in the universe) the widely accepted cosmological model was the standard hot big bang model [62], which describes in detail the evolution of the Universe from a fraction of a second after the beginning, when it was just a hot soup of elementary particles to the present some $13.7 \times 10^9$ years later, when it is filled with stars, planets, galaxies, clusters of galaxies and us [4].

There were also many alternative models, among them the Singularity-Free Cosmological Model [61], which describes an oscillatory Universe that underwent a contraction phase before the present expansion. After a prematter period and a transition period this model at $t = 1.12 \times 10^{-38}$ sec entered a radiation period and became essentially the standard model: the behavior was as if it had begun with the big bang.

DM and cosmic acceleration are verified observationally. It is believed that DM contributes ~24%, and DE up to 71% to the present mass/energy content of the Universe. DE affects the expansion rate of the Universe, which in turn affects the growth of structure and the distances to objects.

In the present scenario, which is carried out in the framework of the Weyl-Dirac version of Wesson's IMT with Rosen's approach, a warped, but empty from matter (cf. Appendix) 5D bulk induces all forms of matter in the 4D brane.

In sections 2 and 3 the geometric background is presented and the EQ.-s in the bulk are derived from a 5D action integral. As dynamical variables serve the metric tensor $g_{AB}(x^C)$, the Weyl connection vector $\tilde{w}_C(x^A)$ and the Dirac gauge function $\Omega(x^A)$. It was proved that the $\Omega$-equation is a corollary of the $g$-equation and of the $\tilde{w}$-equation, so that $\Omega$ may be chosen arbitrarily.

In Sect. 4 field equations on the 4D brane (our Universe) are derived. Having in mind cosmological applications at large distances in a homogeneous and isotropic Universe, we divided the Weyl vector into a global gradient vector $w_{\nu\,glob} = -\dfrac{\Omega_\nu}{\Omega}$ and a chaotically oriented local vector field $w_{\nu\,loc}$ (cf. (24)), which acts inside very small 4D micro-regions, cells.

The local fields are considered in Sec. 5. It turns out that inside micro-cells may be produced particles having mass and spin 1, weylons. As the vector fields inside the cells are chaotically oriented, the summary angular momentum in a macroscopic region vanishes, and there is only a mass effect caused by weylons, so that one can consider the ensemble of weylons as dark matter. One should distinguish between the DM in and around cosmic inhomogeneities (like galaxies, clusters, etc.) and the DM pervading all of the cosmic space. The first one, known as Cold Dark Matter (CDM), may be presented by the ensemble of weylons.

Weylons created by the chaotically oriented fields in micro cells cannot be detected directly as they interact only gravitationally with conventional matter. But one can perhaps estimate roughly the weylons mass from flat rotation curves of spiral galaxies. From the rotation curves as well from massive bodies located in the galaxies' center (supposed they are made of weylons) one can estimate the account of weylon CDM in the Universe.

In Sec. 6 we considered the global gravitational field; its total energy-momentum tensor may be split into three parts: the first presenting the induced by the bulk conventional, luminous matter, the second giving the contribution of the CDM formed by weylons, the third part is formed from the 4D Dirac gauge function $\beta(t)$ and, presumably after an appropriate choice of $\beta$, it will create DE and global DM.

Finally, in the Appendix it is shown that the bulk is a warped, but empty from matter one.

In order to obtain a cosmological model with luminous matter, DM and DE the following procedure may be proposed.

**1**. One starts with EQ.-s (41-46), and setting $\rho_{CDM} = P_{CDM} = 0$, $\beta = 1$, he considers only conventional matter, that at present make up ~4% of the mass/energy content of the Universe. With observational data for luminous matter he construct a skeleton model and chooses an appropriate function $\varphi(t)$; the latter regarded as the creator of conventional matter.

**2**. Taking the observational data for DM, one adds weylon CDM to the skeleton model according to Sec. 5 and to [57, 58].

**3.** Finally with observational data for DE and cosmic acceleration and guided by EQ.-s (41), (42), (45) one construct an appropriate Dirac gauge function $\beta(t)$. It must be recalled that the 4D Dirac gauge function $\beta(t)$ is an arbitrary positive time-depending function in the homogeneous and isotropic Universe, so that the dark matter given by EQ. (45) is pervading all of the cosmic space. This is a Global DM effect and one must distinguish between this Global DM and Cold DM, which is invoked by weylons and located in and around cosmic inhomogeneities, like stars, galaxies, clusters etc.

A paper based on the above-considered procedure with recent cosmological data will be presented in the near future.

APPENDIX

Above we have considered a framework of inducing matter by the 5D bulk in the 4D brane. We obtained luminous matter, DM and DE, the latter being regarded as causing the expansion of the Universe to speed up, The procedure is carried out in the framework of Wesson's IMT, in its Weyl-Dirac modification, with Rosen's approach.

We recall that at large scale the bulk was regarded as isotropic and homogeneous, so that no cosmic (global) vector functions can be present. However, we adopted the existence of microscopic areas, cells, with chaotically oriented Weyl vector fields $\tilde{w}_{A \, \text{loc}}$. The latter under certain circumstances might create massive particles. As in induced matter theories the bulk has to be a pure geometric one, we show below that in our framework the bulk can be made empty from matter by plausible conditions.

It was found that one can describe both, local and global fields by the separation condition (18)

$$\tilde{w}_A = \tilde{w}_{A \, \text{loc}} + \tilde{w}_{A \, \text{global}} = \tilde{w}_{A \, \text{loc}} - \frac{\Omega_A}{\Omega} \qquad (A\text{-}1)$$

Let us consider the local field. It exists in very small micro-cells and is given by EQ. (20):

$$\left( \Omega \tilde{W}^{AB}_{\text{loc}} \right)_{:B} = \frac{1}{2} \sigma \Omega^3 \, \tilde{w}^A_{\text{loc}} \qquad (A\text{-}2)$$

Discarding the subscript $_{\text{loc}}$ and taking into account the smallness of the region, so that for the Dirac gauge function one can take its value $\Omega(c)$ at the center of the cell, we rewrite

$$\left( \tilde{W}^{AB} \right)_{:B} = \frac{1}{2} \sigma \Omega^2(c) \tilde{w}^A \qquad (A\text{-}3)$$

This is the covariant form of the Proca equation [56] for a vector boson field, however in 5 dimensions. From (A-3) follows an a' la Lorentz condition

$$\tilde{w}^A_{:A} = 0 \tag{A-4}$$

Expressing (A-3) in terms of $\tilde{w}^A$, making use of (A-4) and neglecting inside the cell the term with the curvature we obtain.

$$g^{BC}\tilde{w}^A_{:B:C} = \frac{1}{2}\sigma\Omega^2\tilde{w}^A \tag{A-5}$$

Inside the 5D cell, we take the line element as

$$dS^2 = c^2 dt^2 - (dx_1^2 + dx_2^2 + dx_3^2) + \varepsilon\Phi^2 dl^2 \tag{A-6}$$

Let us locate the center of the cell on the brane $\Sigma_0(l_0)$ and denote by $\Phi(l_0) \equiv \Phi_0 = Const$ the value of the lap function in the center of the cell. Then from (A-5) follows

$$\frac{\partial^2 \tilde{w}^A}{c^2 \partial t^2} - \nabla^2 \tilde{w}^A + \varepsilon\Phi_0^2 \frac{\partial^2 \tilde{w}^A}{\partial l^2} = \frac{1}{2}\sigma\Omega^2 \tilde{w}^A \tag{A-7}$$

We assume that in $\tilde{w}^A$ one can separate the dependence on the fifth coordinate from that on the space-time coordinates, so that

$$\tilde{w}^A(t, x_1, x_2, x_3, l) = \Psi(l) \cdot w^A(t, x_1, x_2, x_3) \tag{A-8}$$

Then (A-7) may be rewritten as

$$\Psi\left(\frac{\partial^2}{\partial t^2} - \nabla^2\right) w^A(y^\lambda) = \left(\frac{1}{2}\sigma\Omega^2 \Psi(l) - \varepsilon\Phi_0^2 \frac{d^2\Psi(l)}{dl^2}\right) w^A(y^\lambda); \quad (y^\lambda \equiv t, x^1, x^2, x^3) \tag{A-9}$$

Generally, an observer placed on the brane will recognize (A-9) as an equation describing a massive particle. However provided [4]

$$\Psi''(l) = \frac{1}{2}\frac{\sigma\Omega^2}{\varepsilon\Phi_0^2}\Psi(l) \tag{A-10}$$

one obtains from EQ. (A-8) the wave equation

---

[4] EQ. (A-10) has the simple solutions

For $\lambda \equiv \frac{1}{2}\frac{\sigma\Omega^2(l_0)}{\varepsilon\Phi_0^2} = Const > 0$; one has $\Psi(l) = C_1 \cosh(l\sqrt{\lambda}) + C_2 \sinh(l\sqrt{\lambda})$.

For $\lambda \equiv \frac{1}{2}\frac{\sigma\Omega^2(l_0)}{\varepsilon\Phi_0^2} = Const < 0$; one has $\Psi(l) = C_1 \cos(l\sqrt{|\lambda|}) + C_2 \sin(l\sqrt{|\lambda|})$.

$$\left(\frac{\partial^2}{\partial t^2} - \nabla^2\right) w^A\left(y^\lambda\right) = 0 \qquad (A\text{-}11)$$

Conditions (A-8) and (A-10) will be imposed on the local chaotically oriented Weyl vector fields, so that in the bulk can exist gravitational and Weyl waves (A-11), but no matter. The bulk is warped and empty. It must be pointed out that imposing (A-8, A-9, and A-10) does not affect or restrict the discussion and results presented in the present work.